\documentclass[review]{elsarticle}

\usepackage[T1]{fontenc}
\usepackage{amsmath,amsfonts,amssymb}
\usepackage{graphicx}
\usepackage{algorithm}
\usepackage{algorithmic}
\usepackage{booktabs}
\usepackage{tabularx}
\usepackage[hidelinks]{hyperref}

\journal{Computers in Biology and Medicine}

\begin{document}

\begin{frontmatter}

\title{Reinforcement Learning for Antibiotic Stewardship in a Simulated Hospital Ward with Coupled Within-Host and Transmission Dynamics}

\author[lehigh]{Hsuan-Wei Lee\corref{cor1}}
\ead{hsl324@lehigh.edu}

\author[ntust]{Te-Chun Liu}
\ead{m11209129@mail.ntust.edu.tw}
\cortext[cor1]{Corresponding author}

\affiliation[lehigh]{organization={College of Health, Lehigh University},
            city={Bethlehem},
            state={PA},
            country={USA}}

\affiliation[ntust]{organization={Department of Information Management, National Taiwan University of Science and Technology},
            city={Taipei},
            country={Taiwan}}

\begin{abstract}
Antimicrobial resistance is driven in part by how antibiotics are prescribed, which makes stewardship a central but difficult task. Each prescribing decision affects both the patient treated now and the resistance faced by the whole ward later. We study this trade-off with reinforcement learning. We build a virtual hospital ward that couples a dual-channel transmission model to a within-host competition between susceptible and resistant bacteria, where resistant bacteria carry a fitness cost. A prescriber agent, trained with Proximal Policy Optimization, chooses among antibiotics of different kill rate, cost, and residual activity against resistance, and optimizes a multi-objective reward that places treatment efficacy, resistance control, cost, and adverse effects on a common scale. The agent learns a severity-dependent, tiered strategy without a fixed severity-to-drug rule, although the reward's severity-dependent terms favor this pattern. The policy showed severity-dependent action preferences, consistent with the model's severity-dependent reward and treatment-effect assumptions, and rarely selected the intermediate drug. In 50 evaluation episodes of a 200-slot, 30-day simulation, the PPO policy had a lower mean terminal ward-averaged resistant fraction than random prescribing (48.24\% versus 62.63\%) and a higher model-defined cure rate among ever-infected patients (48.18\% versus 32.85\%). The broad-spectrum drug was selected more often under PPO than under random prescribing. The policy trained at the highest resistance penalty had a higher resistant fraction and a lower cure rate than the others. These results describe the specified simulation and have not been validated against hospital data.
\end{abstract}


\end{frontmatter}

\section{Introduction}
\label{sec:intro}

Antimicrobial resistance (AMR) is now one of the clearest threats to modern medicine. A systematic analysis estimated that bacterial AMR was associated with 4.95 million deaths in 2019, of which 1.27 million were directly attributable to resistance \cite{murray2022burden}. Later work extended these estimates across three decades and projected a rising toll through 2050 \cite{gbd2024amr}. Much of this burden is shaped inside hospitals, where vulnerable patients, dense contact, and heavy antibiotic use let resistant organisms spread and persist. Antimicrobial stewardship, the effort to use antibiotics only when they help and to prefer narrower agents where possible, is the main lever clinicians have against this process. The difficulty is that stewardship asks a prescriber to weigh a concrete benefit for one patient against a diffuse and delayed cost to everyone treated later.

That tension is a collective-action problem. When each prescriber does what is best for the patient in front of them, the shared resource of antibiotic effectiveness erodes for the whole ward. Porco and colleagues framed antibiotic overuse as a tragedy of the commons and showed, using transmission models analyzed as a game, that private and social interests need not align \cite{porco2012commons}. A game-theoretic study of prescribing reached a similar conclusion, namely that rational doctors keep prescribing regardless of the resistance they help create, unless the payoffs they face are changed \cite{colman2019game}. This framing shapes our choice of method. If the core problem is a misalignment of incentives, then a learning agent that optimizes an explicit, shaped reward is a natural way to study which incentives steer behavior toward better collective outcomes. We have used multi-agent learning in this spirit before, showing that adaptive agents can raise collective welfare in a social dilemma \cite{lee2025granular}.

Stewardship decisions do not act on resistance directly. They act through two coupled processes. The first is transmission within the ward, where resistant strains move between patients by direct contact and through a contaminated environment. Ward-level models have been built to study this spread and to test infection control and stewardship interventions \cite{durazzi2023amr, cheng2024nosocomial}. The second is the within-host competition between susceptible and resistant bacteria. Resistance usually carries a fitness cost, so a resistant strain grows more slowly when the drug is absent, which is what makes reduced use able, in principle, to recover susceptibility \cite{andersson2010cost}. The size and timing of that effect depend on dose, drug action, and competition inside the host \cite{czuppon2023stochastic}. We therefore couple a transmission layer to a within-host layer, rather than treat resistance as a single aggregate number.

Reinforcement learning (RL) has become a strong candidate for sequential clinical decisions. The AI Clinician learned fluid and vasopressor strategies for sepsis from intensive care records, and recommended choices that, when matched by clinicians, were associated with lower mortality \cite{komorowski2018clinician}. Related work has learned dynamic treatment regimes from observational data \cite{abebe2024dtr} and has incorporated clinical guidelines into reinforcement-learning frameworks to guide treatment recommendations, for antibiotic therapy \cite{wang2024antibiotic} and for mechanical ventilation \cite{denhengst2024ventilation}. For resistance in particular, RL has been used to sequence drugs so that bacterial populations are steered toward collateral sensitivity \cite{weaver2024resistance}. Machine learning has also been used to predict resistance from hospital time series, which shows the signal is learnable but stops short of acting on the stewardship trade-off \cite{hernandez2023amr}. Earlier reinforcement learning for resistance has mostly treated a single patient or a fixed environment, rather than a ward whose resistance evolves in response to prescribing. We examine prescribing decisions in a simulated ward that couples within-host bacterial dynamics with between-patient transmission, so that the long-run externality of each choice becomes part of what the agent learns. Our own recent work took a first step in this direction for epidemic control, using spatial multi-agent reinforcement learning with heterogeneous risk preferences \cite{liu2027spatial}. The present study moves that idea into antibiotic stewardship.

We build a virtual hospital ward in which a prescriber agent, trained with Proximal Policy Optimization \cite{schulman2017ppo}, chooses among antibiotics of different kill rate, cost, and residual activity against resistance. The environment couples a dual-channel transmission model to a within-host growth model with mutation and a fitness cost for resistance. The agent optimizes a multi-objective reward that places treatment efficacy, resistance control, cost, and adverse effects on a common scale, which internalizes the stewardship externality instead of encoding a fixed prescribing rule \cite{roijers2013morl}. We summarize the learned behavior and compare the policy against rule-based baselines, a comparison for which a Pareto view of health and cost trade-offs has been used before \cite{janko2023npi, wulkow2021pareto}. The framework links patient-level bacterial responses to ward-level transmission, so a prescribing policy can be evaluated against both a treatment outcome and a ward-level resistance measure.

First, the agent learns a severity-dependent, tiered prescribing pattern without a fixed severity-to-drug rule, selecting the broad-spectrum drug frequently for severe cases and the narrow-spectrum drug frequently for mild cases, and rarely the intermediate one. Second, relative to random prescribing it lowers the ward-averaged resistant fraction while raising the cure rate, and within an episode its use of the broad-spectrum drug falls over time, even though its overall broad-spectrum proportion stays higher than under random prescribing. Third, among policies trained at different resistance penalties, the one at the highest penalty had a higher resistant fraction than the others, which we treat as an observation from the model rather than a demonstrated mechanism.

\section{Methods}
\label{sec:meth}
\subsection{Virtual hospital ward}
We represent the clinical setting as a virtual hospital ward, modeled as a two-dimensional spatial grid that stands in for the complex pattern of interpersonal patient contact~\cite{kerkmann2025abm}. The ward measures $L \times L$ meters and holds up to $N$ patients, with $N = 200$ on an $L = 50$ m grid in all experiments. The default environmental, epidemiological, and reward parameters used throughout are listed in \ref{app:params}.

\subsection{Transmission dynamics}
Resistant strains spread within the ward through two channels~\cite{durazzi2023amr, cheng2024nosocomial, changruenngam2022crkp}. The model combines an environmental channel ($\beta_{env}$) and a contact channel ($\beta_{contact}$). Infected patients shed bacteria that diffuse across the ward through a spatial Gaussian kernel. Direct interpersonal transmission is treated as a discrete probabilistic event within a contact radius $r_c = 2$ m, depending on the distance between patients and on the resistance ratio of the source patient, given by:
\begin{equation}
P_{contact} = 1 - \exp(-\beta_{contact} \cdot \rho_{R, source} \cdot \mathbb{I}(d < r_c))
\end{equation}

\subsection{Observation space}
The observation for each prescriber is a continuous 12-dimensional vector of type \texttt{float32} that combines patient-specific attributes with a partial view of the ward. The first two entries are the normalized coordinates $(\frac{x}{L}, \frac{y}{L})$. The third is the infection severity, coded as $\frac{1}{3}$, $\frac{2}{3}$, or $1$ for mild, moderate, and severe cases. The fourth is the ward-level mean resistant fraction, averaged over all patients. The fifth is the normalized timestep $\frac{t}{T}$. The sixth to eighth entries are three binary comorbidity indicators, assigned at random when a patient is admitted and held fixed for the episode. They do not influence the within-host dynamics, the transmission, or the reward, so in the current setup they carry no information the policy can use, and we keep them as placeholders for a future comorbidity extension. The ninth entry is the drug prescribed to the patient on the previous day. The tenth is the change in bacterial load relative to the previous day, on a logarithmic scale. The eleventh is a lagged and noisy estimate of the local resistant-bacteria density around the patient, where the added noise represents each prescriber's imperfect awareness of nearby resistance. The twelfth is the current total bacterial load, as $\frac{\log_{10}(\text{load})}{9}$, which lets the agent respond to infection dynamics that span several orders of magnitude. Comorbidities therefore enter only as these uninformative placeholder indicators in the current model.

\subsection{Within-host dynamics and treatment effect}
Within each host, the infection evolves through a pair of deterministic growth equations. The susceptible ($S$) and resistant ($R$) populations grow logistically and compete for a shared carrying capacity $K$, so within-host selection depends on drug exposure rather than on any fixed rule. Mutation enters as a deterministic flux at rate $m$ from the susceptible to the resistant population. We assume a fitness cost of resistance, represented by a lower intrinsic growth rate $r_R$ for the resistant population than the susceptible rate $r_S$, which reflects the cost of resistance when the drug is absent~\cite{andersson2010cost, davies2019withinhost, czuppon2023stochastic}.

\begin{equation}
\frac{dS}{dt} = r_S S \left(1 - \frac{S+R}{K}\right) - \mu_S S - \phi(c)S - mS
\end{equation}
\begin{equation}\frac{dR}{dt} = r_R R \left(1 - \frac{S+R}{K}\right) - \mu_R R - \alpha(c)\phi(c)R + mS
\end{equation}
Here $c$ denotes the antibiotic chosen for the patient, $\phi(c)$ is that drug's kill rate against the susceptible population, and $\alpha(c)$ is its residual efficacy against the resistant population, so $\alpha(c)\phi(c)$ is the kill rate against resistant bacteria. These equations are integrated with a forward Euler scheme, using ten substeps per simulated day ($\Delta t = 0.1$ day), which keeps the update numerically stable.

Each day the within-host update is applied to patients whose load is above $10^4$ CFU, which we treat as the threshold for active infection. A patient whose load falls between $10^3$ and $10^4$ CFU is no longer updated unless a transmission event raises it above $10^4$ CFU again. The simulator can remove a patient on recovery, when the load falls below $10^3$ CFU, or on treatment failure, when it reaches $0.9 \times 10^9$ CFU, but no such removal occurs in the recorded episodes, so the cohort stays at $N = 200$ throughout. These thresholds are operational choices for the simulation and are not validated clinical criteria. For learning, the efficacy reward instead uses a continuous logarithmic reduction, $\log_{10}(pre\_load) - \log_{10}(post\_load)$, which gives the dense gradient signal that policy optimization needs.

Antibiotic use adds selection pressure that reshapes bacterial growth within the host. Each drug enters the growth equations through its kill rate $\phi(c)$ against the susceptible population and its residual efficacy $\alpha(c)$ against the resistant population, both as subtractive terms. Clearance therefore depends on which drug is chosen, through its $\phi(c)$ and $\alpha(c)$. As one exception, when a mild (Severity 1) patient receives broad-spectrum Drug 2 the kill rate $\phi$ is halved, which weakens an inappropriate broad-spectrum choice on top of the reward penalty described below.

\subsection{Reinforcement learning agent}
We train the prescriber agent with Proximal Policy Optimization (PPO)~\cite{schulman2017ppo}. A single policy network is learned and shared across all prescribers, each of whom applies it to their own patient. The environment instantiates three prescriber profiles with different reward weight vectors $(\omega_\alpha, \omega_\beta, \omega_\gamma, \omega_\delta)$: a guideline profile $(1.9, 1.8, 0.3, 0.1)$ for 50\% of prescribers, a conservative profile $(1.8, 2.0, 0.3, 0.1)$ for 30\%, and an aggressive profile $(2.0, 0.2, 0.2, 0.2)$ for 20\%. The four weights apply to treatment efficacy ($\omega_\alpha$), resistance prevention ($\omega_\beta$), financial cost ($\omega_\gamma$), and adverse-outcome penalties ($\omega_\delta$). The profiles differ mainly in how they weight resistance against efficacy, with the conservative profile weighting resistance most and the aggressive profile least. The shared policy is trained on rewards generated under all three profiles. Because profile identity is not part of the observation space, the policy cannot condition its actions on prescriber type. Prescribing showed only a weak association with profile (Cramer's V $< 0.1$), so we analyze the shared policy as a single learned strategy. Because the reward components are measured on different numerical scales, the weights alone should not be read as a direct ranking of their importance. Adding prescriber characteristics to the observation space, so that heterogeneous behavior can be represented and separated, is a clear direction for future work.

\subsection{Action space}
The action space has three discrete antibiotic choices. These are hypothetical treatment options rather than specific agents, and they differ in kill rate, cost, and residual activity against the resistant population. We use the narrow-spectrum and broad-spectrum labels only as a shorthand, since the model does not track which pathogens a drug covers. Keeping three options keeps the ward-level trade-off between efficacy, resistance control, and cost visible, and it lets us study the core prescribing decision without a high-dimensional action space. Drug 0 has a kill rate of $\phi_0 = 2.2$ against the susceptible population and the lowest cost of $0.1$. Drug 1 has an intermediate susceptible kill rate of $\phi_1 = 2.4$ and a cost of $1.0$. Drug 2 has the highest susceptible kill rate ($\phi_2 = 2.5$) and the highest cost ($3.0$), and its overuse draws severity-dependent prescribing penalties in the reward. The residual-efficacy factors ($\alpha_0, \alpha_1, \alpha_2$) in \ref{app:params} set how well each drug acts against the resistant population, and they do not rise monotonically with the susceptible kill rate.

\subsection{Multi-objective reward}
The agent is trained on a multi-objective reward $R_{i}^{t}$~\cite{roijers2013morl} that balances short-term outcomes for the individual patient against long-term ward-level resistance. The Treatment Efficacy term rewards bacterial clearance and load reduction on a logarithmic scale. The Adverse Effects term applies a small fixed penalty when a patient is not cured, as a stylized stand-in for treatment-related harm. The model does not simulate specific adverse events such as drug toxicity or Clostridioides difficile infection. A set of severity-dependent prescribing terms then shapes behavior. They reward broad-spectrum use for the most severe patients and penalize it for mild patients, with the exact values given in the implementation details. These terms are added to the Treatment Efficacy term before it is weighted, so they do not appear as a separate term in the consolidated reward below.

An economic cost term reflects daily drug use and the larger financial burden of treatment failure, which consumes extra care resources. The resistance term is a continuous quadratic penalty that combines each patient's own resistant fraction with the ward-level mean. Together these terms internalize the stewardship externality, so that lowering resistance competes with immediate efficacy during training.

\begin{equation}
R_{resist} = \eta ( \rho_{i} + \omega \rho_{p}^{2} )
\end{equation}
Here $\rho_i = R_i/(S_i + R_i)$ is patient $i$'s own resistant fraction, and $\rho_p$ is the ward-level mean of $\rho_i$ taken over all patients. The scalar $\eta = 0.4$ sets the overall strength of the resistance penalty, and $\omega = 20$ weights the squared ward-level term relative to the patient's own fraction.

At the end of each environment step, the reward components are combined into a single scalar for policy optimization. The consolidated reward $R_{i}^{t}$ is:
\begin{equation}
R_{i}^{t} = \omega_\alpha R_{treat} - \omega_\beta R_{resist} - \omega_\gamma R_{cost} - \omega_\delta R_{adverse}
\end{equation}
The efficacy term $R_{treat}$ is added, while the resistance, cost, and adverse-effect terms are defined as non-negative penalties and subtracted.

\subsection{Evaluation metrics}
We evaluate the trained policy on treatment outcomes and resistance control. The two primary measures are the model-defined cure rate and the ward-averaged resistant fraction. The model-defined cure rate is the proportion of patients who were actively infected at some point in the episode and whose load then fell below $10^4$ CFU at least once during that episode. A patient is counted as cured at the first such drop, and that status is not revoked if the load later rises again, so the measure reflects whether an infection was ever brought under control rather than its state on the final day. The denominator is this ever-infected group, so the cure rate is read over infected patients rather than over the whole ward. The resistant fraction is the arithmetic mean over all $N = 200$ patient-slots of $\rho_i = R_i/(S_i+R_i)$, with the denominator set to one when a patient carries no bacteria, reported as a percentage. Patients who carry no bacteria contribute zero, so the average runs over the whole ward rather than over infected patients only. The measure therefore depends both on the resistant fractions of patients carrying bacteria and on how many slots are zero. A change in it does not necessarily indicate a change in the total resistant bacterial burden. We report it either daily along the trajectory or as a terminal value at the final timestep $t = T = 30$ days. We then compare the drug-usage distribution of the RL agent over time against a random baseline prescriber.

\subsection{Baseline policies for comparison}
To benchmark the PPO agent against reference strategies, we add two deterministic rule-based policies alongside the unguided random baseline.
\begin{itemize}
    \item \textbf{Severity-based baseline:} This policy matches the antibiotic to the patient's infection severity. Mild cases (Severity 1) receive narrow-spectrum Drug 0, moderate cases (Severity 2) receive intermediate-spectrum Drug 1, and severe cases (Severity 3) receive broad-spectrum Drug 2. This is a simplified simulation comparator and not an implementation of a specific clinical guideline.
    \item \textbf{Drug 2 restriction baseline:} Designed to minimize selection pressure, this policy bans broad-spectrum Drug 2 entirely. Severe cases (Severity 3) are limited to intermediate-spectrum Drug 1, while mild and moderate cases (Severity 1 and 2) receive only narrow-spectrum Drug 0.
\end{itemize}
These static, severity-based policies give fixed reference points, against which we test whether a dynamic, environment-aware agent can do better.

\subsection{Implementation and training details}
The ward is a closed cohort of $N = 200$ patients, with no admissions during the 30-day episode. At initialization, 20\% of patients are infected, half with a susceptible-dominant load ($10^5$ susceptible and $10^3$ resistant CFU) and half with a resistant-dominant load ($10^3$ and $10^5$). The environmental grid is rebuilt each day and carries no multi-day decay. Infected patients shed resistant bacteria onto the grid at a fraction $\omega_{env} = 10^{-4}$ through the Gaussian kernel. A successful environmental transmission adds $10^3$ resistant CFU to a patient, and a successful contact transmission adds $10^4$ CFU split between the susceptible and resistant pools in proportion to the source patient's own mix. The severity-dependent shaping adds $+15$ to the reward when a severe (Severity 3) patient receives Drug 2 and $+5.5$ when a moderate (Severity 2) patient receives Drug 2, and subtracts $80$ when a mild (Severity 1) patient receives Drug 2.

The policy is a multilayer perceptron with two hidden layers of 64 units and tanh activations, with separate actor and critic networks. Training used a learning rate of $5\times10^{-5}$, an entropy coefficient of $0.02$, a clipping range of $0.1$, and eight parallel environments. The main policy in Table~\ref{tab:stewardship_outcomes} was fine-tuned from an earlier baseline checkpoint for four million steps, and the penalty-setting policies in Figure~\ref{fig:pareto_frontier} were each fine-tuned from the same checkpoint for six million steps. At evaluation, actions are taken deterministically as the arg-max of the policy. We did not fix a global random seed during training, so the training runs are not exactly reproducible, although the saved checkpoints can be re-evaluated with fixed evaluation seeds. The reported outcomes are summarized over 50 evaluation episodes of each fixed trained policy.

\section{Results}

The trained agent holds the ward's terminal resistant fraction at about $48.24\%$, a reduction of $14.39$ percentage points relative to the $62.63\%$ of the random baseline. It reaches this while raising the model-defined cure rate from $32.85\%$ to $48.18\%$, an increase of $15.33$ percentage points. These improvements occur even though the agent selected the broad-spectrum Drug 2 action in $56.61\%$ of cases, more often than the $33.50\%$ under random prescribing. The improvement therefore occurred alongside differences in the overall action mix and in prescribing across severity levels and time, not a reduction in Drug 2 selection overall. A two-sample t-test confirms that the difference in the resistant fraction is statistically significant ($p < 0.001$). Standardized mean differences accompany the hypothesis tests to describe the effect magnitude. Table~\ref{tab:stewardship_outcomes} reports the global outcomes together with the drug-usage distribution.

\begin{table}[htbp]
\caption{{\small Global stewardship outcomes. Resist. (\%) is the ward-averaged resistant fraction, Cure (\%) is the model-defined cure rate among ever-infected patients, and the Drug columns are the policy's action-output frequencies across all patient-slots. \textbf{Bold} marks the better value in each column, and $^{***}$ marks $p < 0.001$. \label{tab:stewardship_outcomes}}}
\begin{center}
\begin{tabular}{|l|c|c|c|c|c|c|}
\hline
\rule[-1ex]{0pt}{3.5ex} Policy & Mean Reward & Resist. (\%) & Cure (\%) & Drug 0 (\%) & Drug 1 (\%) & Drug 2 (\%) \\
\hline
\rule[-1ex]{0pt}{3.5ex} Random & -152.83 & 62.63 & 32.85 & 33.20 & 33.31 & 33.50 \\
\hline
\rule[-1ex]{0pt}{3.5ex} \textbf{PPO (Ours)} & \textbf{-79.22} & \textbf{48.24}$^{***}$ & \textbf{48.18}$^{***}$ & 40.06 & 3.33 & 56.61 \\
\hline
\multicolumn{7}{l}{\rule[0ex]{0pt}{3ex} \footnotesize Average of 50 independent evaluation episodes under the fixed trained policy ($N=200, L=50$).} \\
\multicolumn{7}{l}{\footnotesize P-values are from two-sample t-tests.} \\
\end{tabular}
\end{center}
\end{table}

Across the 50 evaluation episodes, the terminal resistant fraction was $48.24 \pm 5.27\%$ (95\% CI $[46.74, 49.73]$) for PPO and $62.63 \pm 6.29\%$ ($[60.84, 64.41]$) for the random baseline, a standardized difference of $d = -2.48$. The model-defined cure rate was $48.18 \pm 8.18\%$ ($[45.86, 50.51]$) for PPO and $32.85 \pm 6.91\%$ ($[30.89, 34.82]$) for random, with $d = 2.02$. The reported uncertainty reflects variation across evaluation episodes for the fixed trained policy, and it does not quantify variability across independent training runs.

Drug 1, the intermediate option, is almost absent. It accounted for 3.33\% of PPO action outputs, compared with 33.31\% under random prescribing. Under the specified parameters this is understandable. Drug 1 costs more than Drug 0, and its activity against the resistant population is actually lower. Its effective kill rate against resistant bacteria is $\alpha_a \phi_a = 0.192$, below $0.330$ for Drug 0 and $0.500$ for Drug 2, despite its higher kill rate against the susceptible population. These parameter values may help explain why Drug 1 was selected infrequently under the specified reward.

The reward function combines treatment efficacy, a resistance penalty, a cost penalty, an adverse-effect penalty, and the severity-dependent prescribing terms. Reporting these components makes the training objective explicit, though the objective alone does not explain each individual decision made by the policy network.

\begin{figure} [htbp]
    \begin{center}
    \begin{tabular}{ccc}
     \includegraphics[width=0.48\linewidth]{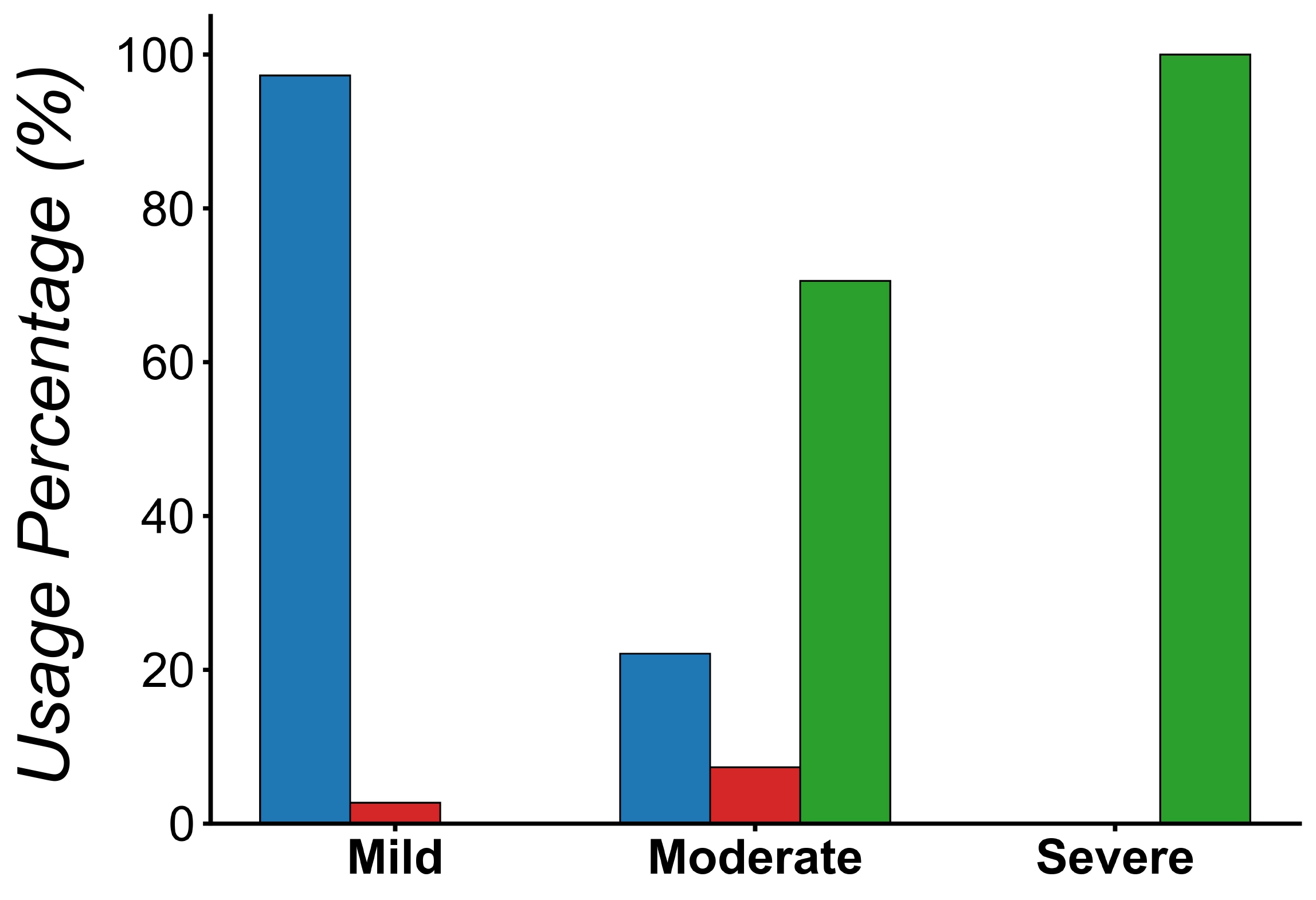} &
     \includegraphics[width=0.48\linewidth]{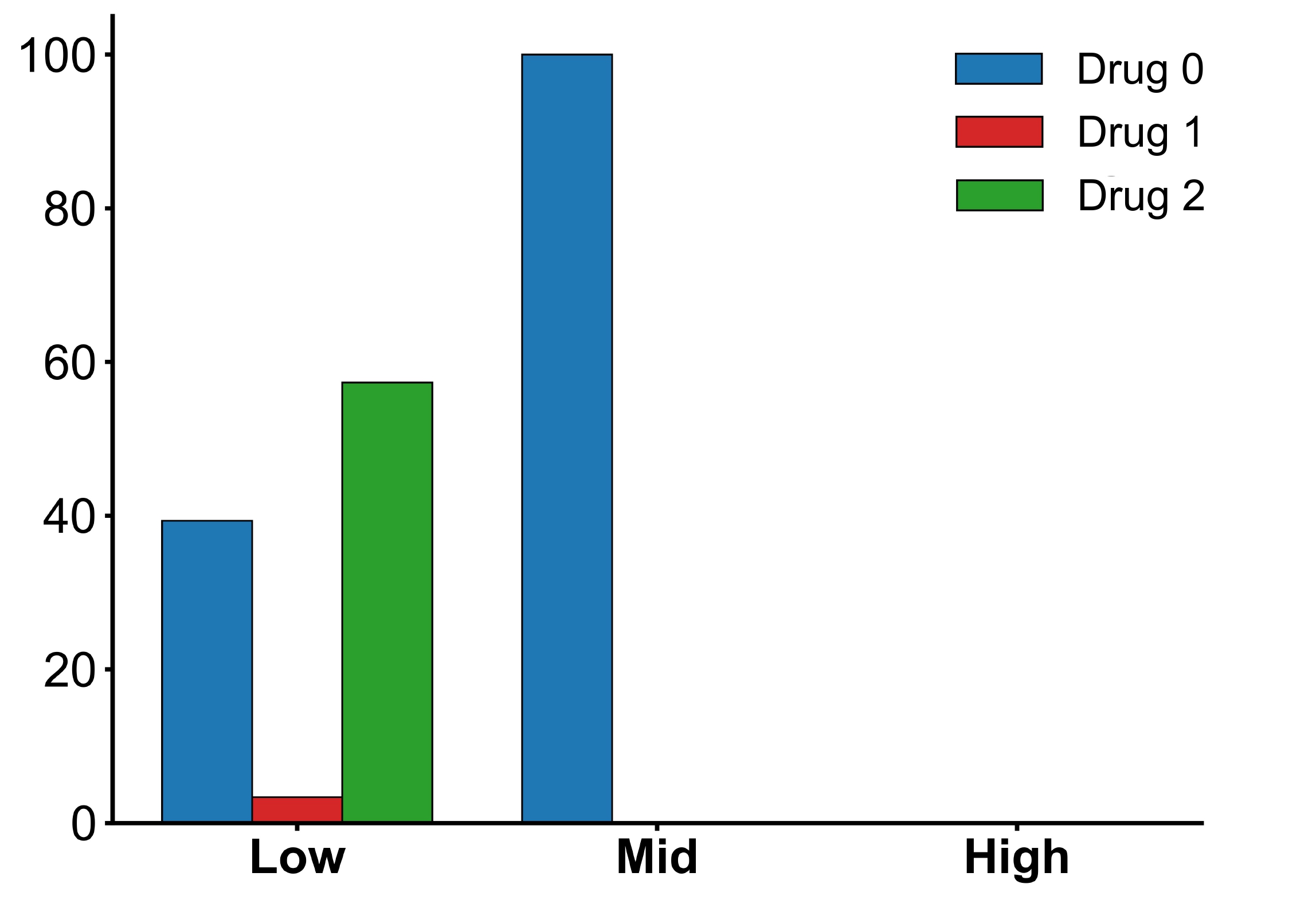} \\
     {\footnotesize (a) Patient Severity} & {\footnotesize (b) Bacterial Load}
    \end{tabular}
    \end{center}
    \caption{{\small Stratified prescribing policy learned by the RL agent. \textbf{Panel (a)} shows the severity gradient, where the most severe cases (Severity 3) receive heavy broad-spectrum use. \textbf{Panel (b)} shows more frequent Drug 2 use at low bacterial load. \label{fig:stratified_policy}}}
\end{figure}

The learned prescribing pattern is severity-dependent, shown in Panel A of Figure~\ref{fig:stratified_policy}. For the most severe patients, the agent gives broad-spectrum Drug 2 almost 100\% of the time. For mild cases at severity 1, it gives narrow-spectrum Drug 0 about 97\% of the time. This pattern is learned without a fixed severity-to-drug rule, although the reward penalizes broad-spectrum use for mild cases and rewards it for severe cases, which favors it.

Bacterial load gives the policy a second axis, shown in Panel B of Figure~\ref{fig:stratified_policy}. Drug 2 was selected more frequently at low bacterial loads. As the load reaches an intermediate range, Drug 0, the narrow-spectrum option, becomes more common. This pattern is consistent with the reward, which rewards early load reduction.

Spatial heterogeneity in resistance density was small in our simulation (coefficient of variation $< 0.15$), which suggests that the environmental channel produces a fairly uniform resistance gradient across the ward. This is why we focus on temporal and severity-based adaptation rather than spatial adaptation. Larger grids, lower diffusion, or more localized transmission would be natural settings in which to probe the limits of spatial stewardship, and we leave these to future work.

\begin{figure} [htbp]
    \begin{center}
    \begin{tabular}{c}
     \includegraphics[width=0.85\linewidth]{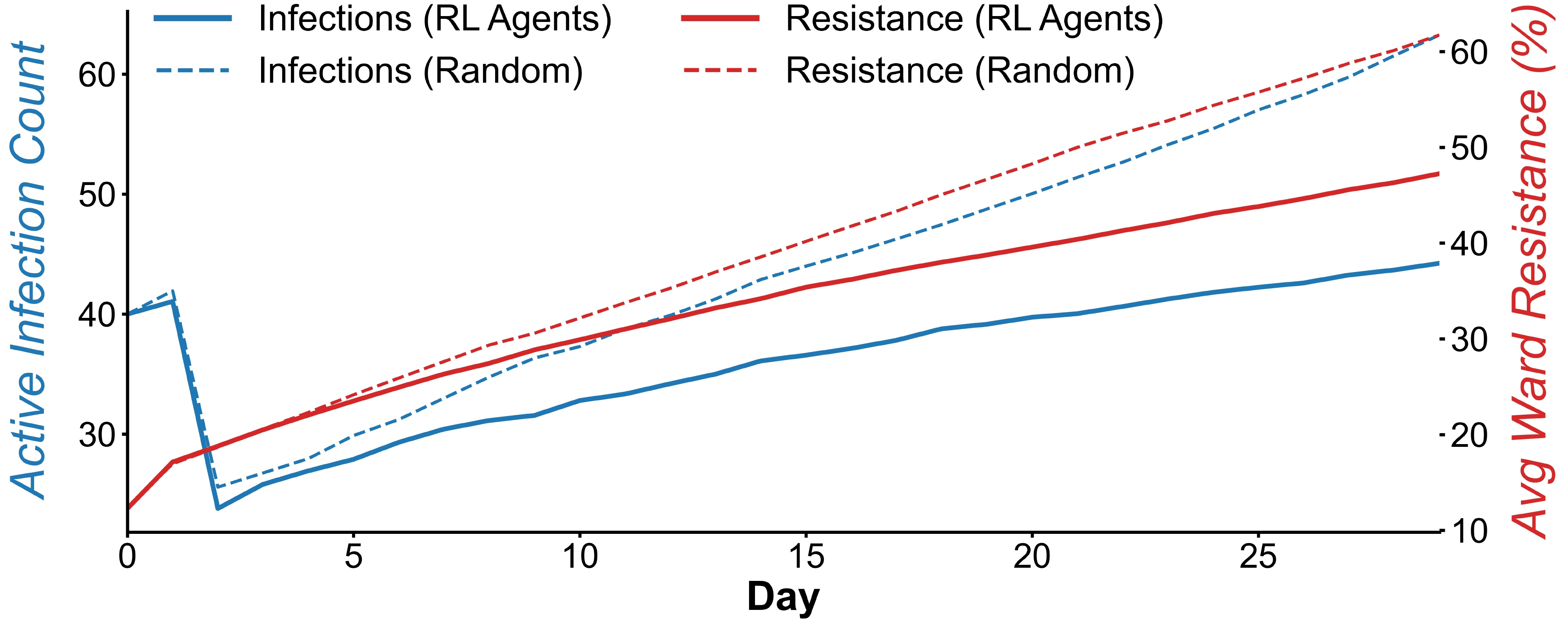} \\
     {\footnotesize (a) Infections vs. Resistance Trajectories (RL vs. Random)} \\[2ex]
     \includegraphics[width=0.85\linewidth]{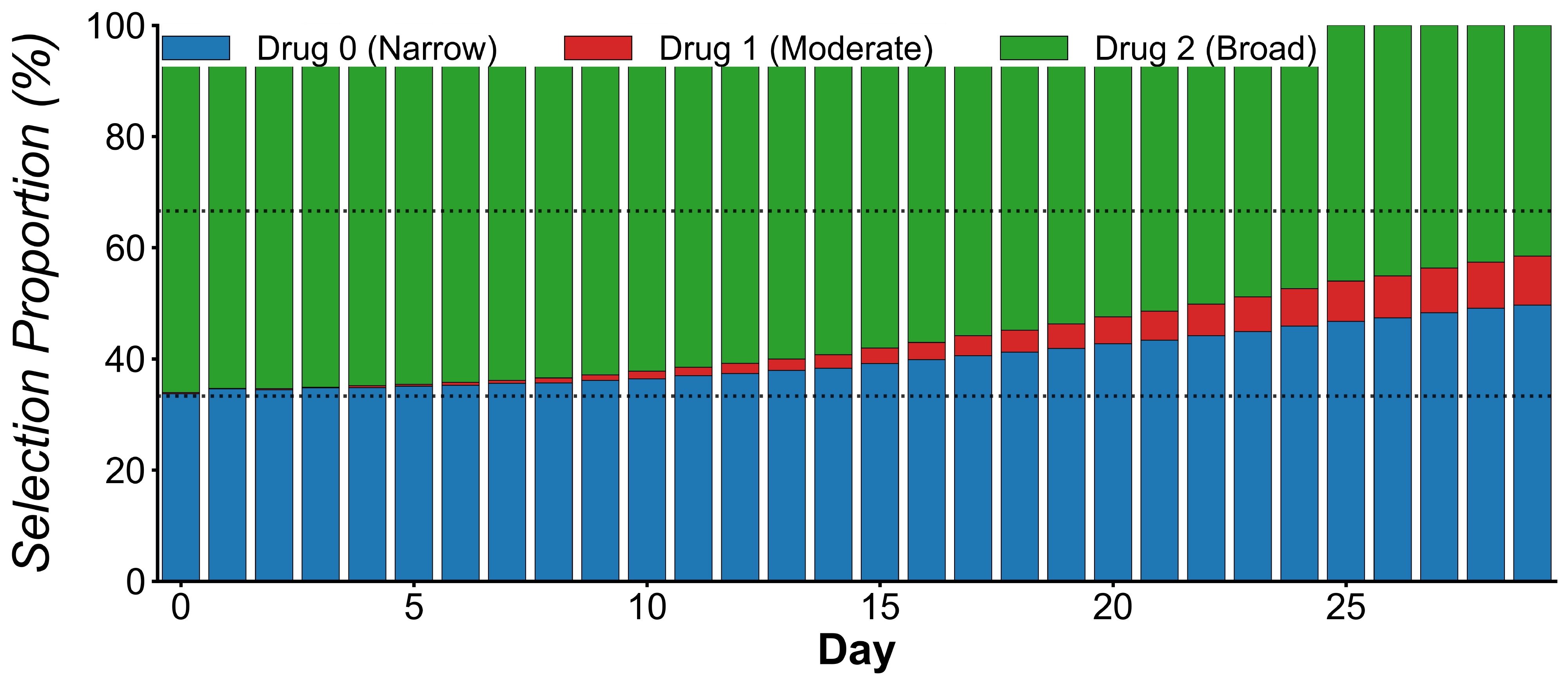} \\
     {\footnotesize (b) Daily Temporal Drug Usage Distribution (PPO Agent)}
    \end{tabular}
    \end{center}
    \caption{{\small Temporal dynamics and baseline comparison. \textbf{Panel (a)} shows the daily resistant-fraction and infection-count trajectories for the PPO agent and the random baseline over a 30-day episode. \textbf{Panel (b)} shows the agent's temporal strategy, which lowers the frequency of the broad-spectrum Drug 2 action from $\sim$67\% to $\sim$38\% over the episode. \label{fig:temporal_dynamics}}}
\end{figure}

The frequency of the Drug 2 action also declined over the 30-day episode, shown in Panel B of Figure~\ref{fig:temporal_dynamics}, dropping from about 67\% on Day 1 to roughly $38$\% by Day 30 as selection pressure built up.

Under PPO, this declining Drug 2 use coincided with a lower resistant-fraction trajectory than under random prescribing, shown in Panel A of Figure~\ref{fig:temporal_dynamics}. The random baseline climbs sharply, while the PPO trajectory settles at a lower level, and the cure rate is higher at the same time.

\begin{figure} [htbp]
    \centering
    \includegraphics[width=0.85\linewidth]{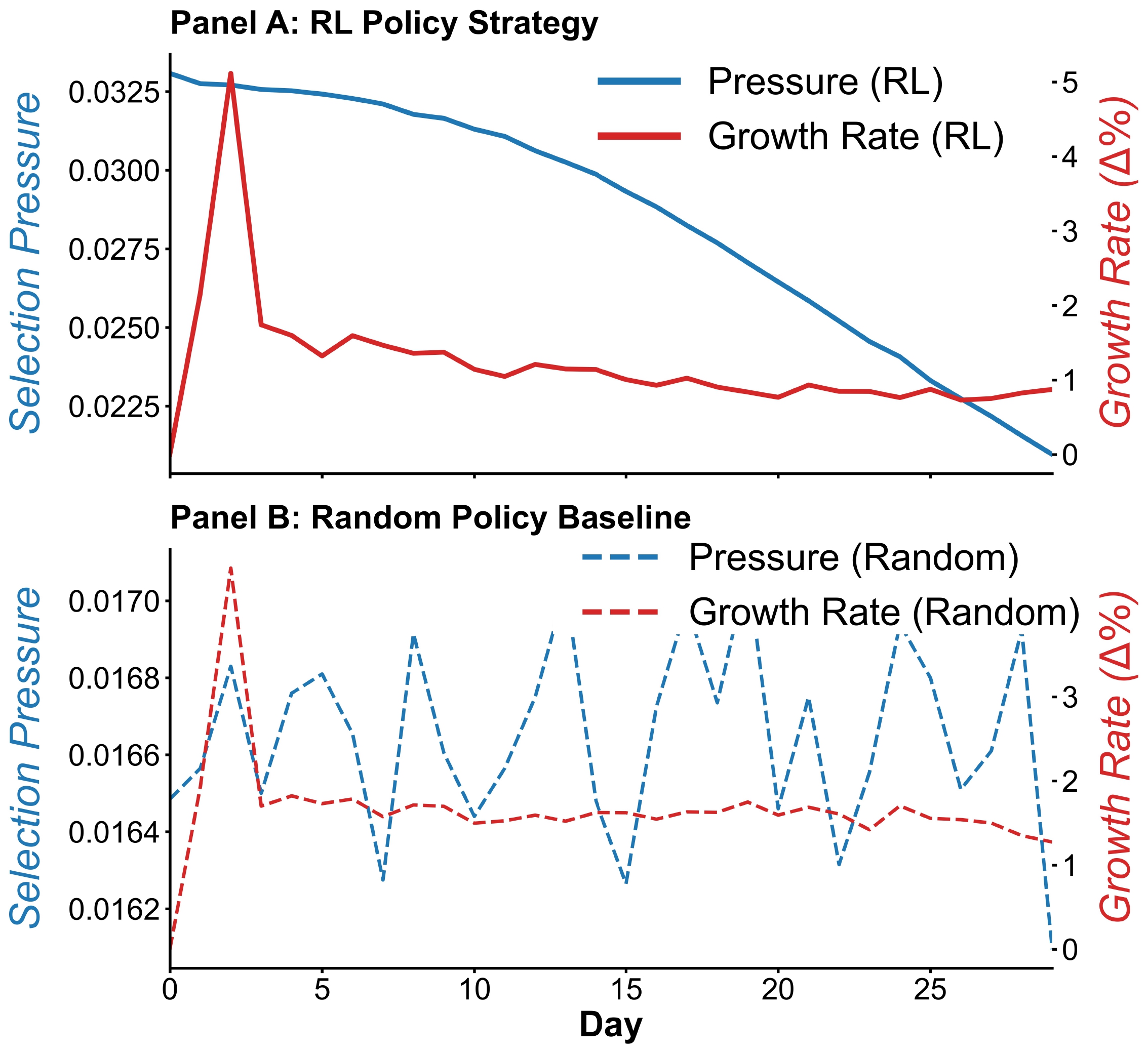}
    \caption{Temporal evolution of the Drug 2 use-based index and resistant-fraction growth, where the index is defined as Drug 2 usage multiplied by the resistance fitness cost. \textbf{Panel (A)} shows that under PPO the index declines after an early peak, and the resistant-fraction growth rate stabilizes. \textbf{Panel (B)} shows the random baseline, which does not suppress the growth rate after Day 3. The two panels use independent vertical axes so that the baseline fluctuations stay visible. \label{fig:F3_comparison}}
\end{figure}

As a further check, we track a Drug 2 use-based index, defined as Drug 2 usage multiplied by the resistance fitness cost. We use it as a simple summary of the ecological pressure from broad-spectrum use in the model, not as a validated measure of selection pressure. The biological role of a fitness cost in the competition between resistant and susceptible strains is well documented~\cite{krishna2025fitness}. Under PPO, the index declined over the episode after an early peak, as shown in Figure~\ref{fig:F3_comparison}. The resistant-fraction growth rate flattens after Day 15. The random baseline, in contrast, shows a resistant fraction that keeps growing and does not stabilize over the simulated period.

\begin{figure} [htbp]
    \centering
    \includegraphics[width=0.95\linewidth]{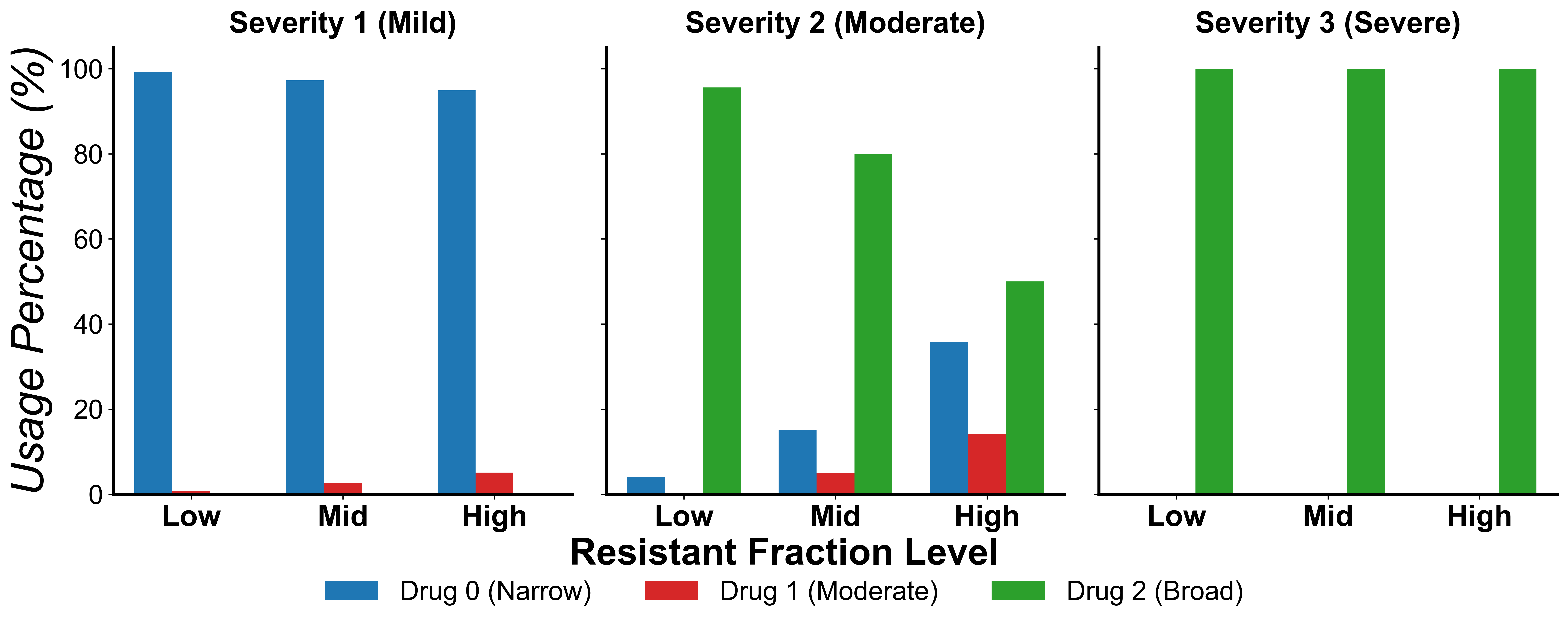}
    \caption{Adaptive policy map. We show the agent's antibiotic selection frequencies across ward-resistance categories (Low, Mid, High) and patient severity (1: Mild, 2: Moderate, 3: Severe). Among slots classified as moderate severity, the policy selected the Drug 2 action in 95.6\% of outputs in the low ward-resistance category and 50.0\% in the high category. The counts are the policy's action outputs across all 200 patient-slots over 30 days and 50 episodes ($200 \times 30 \times 50 = 300{,}000$ in total), which include uninfected and inactive slots: $N_{Low} \approx 81{,}600$, $N_{Mid} \approx 92{,}000$, and $N_{High} \approx 126{,}400$. The high category uses a 35\% threshold. These are descriptive associations. \label{fig:contingent_reasoning}}
\end{figure}

The policy map in Figure~\ref{fig:contingent_reasoning} shows a state-dependent prescribing pattern, clearest among moderate-severity slots. Among slots classified as moderate severity, the policy selected Drug 2 less frequently in the higher ward-resistance category and selected lower and medium-strength drugs more often. For this visualization we record the policy's chosen action for every patient-slot on every day of the 50 evaluation episodes, and group these outputs into ward-resistance categories, with the high category set at a 35\% threshold so that the three categories hold comparable numbers of observations. The outputs include uninfected and inactive slots, not only actively treated patients, so the frequencies describe the policy's action tendency rather than administered prescriptions.

\begin{figure} [htbp]
    \centering
    \includegraphics[width=0.95\linewidth]{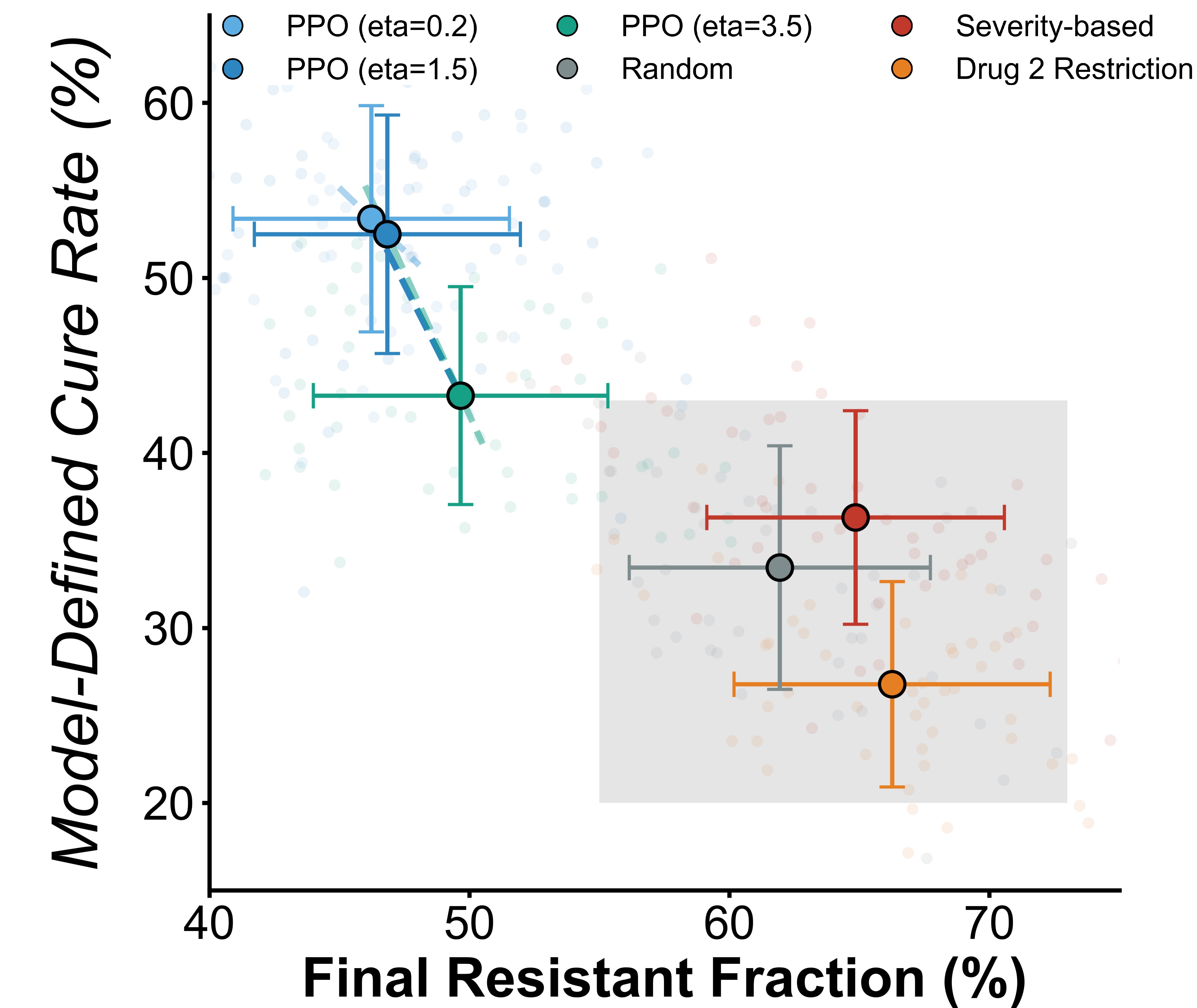}
    \caption{Outcomes of the evaluated prescribing policies. The axes show terminal resistant fraction against the model-defined cure rate, where lower resistance and higher cure indicate better performance. The PPO policies are evaluated at resistance-penalty settings $\eta = 0.2$, $1.5$, and $3.5$. Among the displayed PPO policies, the policy trained at $\eta = 3.5$ had the highest mean terminal resistant fraction and the lowest mean cure rate. The baseline policies (Random, Severity-based, and Drug 2 restriction) had higher mean terminal resistant fractions and lower mean model-defined cure rates than the displayed PPO policies. Markers show the mean across 50 evaluation episodes, error bars show $\pm 1$ standard deviation, and the faint points are individual episode outcomes. Connecting lines serve only as visual guides. \label{fig:pareto_frontier}}
\end{figure}

Figure~\ref{fig:pareto_frontier} compares the outcomes of the evaluated prescribing policies~\cite{reymond2022pareto}. Lower terminal resistant fraction and higher model-defined cure rate indicate better performance. The displayed PPO policies have a lower mean resistant fraction and a higher mean cure rate than the baselines. Among the PPO settings, the policy trained at the highest resistance penalty ($\eta = 3.5$) had a higher mean resistant fraction and a lower mean cure rate than the other displayed PPO policies. These policies are trained under different resistance-penalty settings ($\eta = 0.2, 1.5, 3.5$), separate from the main policy in Table~\ref{tab:stewardship_outcomes}, which uses $\eta = 0.4$. Connecting lines in the figure serve only as visual guides.

To test robustness, we ran a sensitivity analysis with $\pm20\%$ perturbations on $\beta_{env}$, $\beta_{contact}$, and $m$, with full results in Supplementary Table~\ref{tab:sensitivity}. Across the tested perturbations, the mean terminal resistant fraction ranged from $45.95\%$ to $49.70\%$. These comparisons evaluate the same fixed trained policy under perturbed environment parameters, using 10 evaluation episodes per condition.

\section{Discussion}
\label{sec:discussion}

In the main simulation, PPO reduced the ward-averaged resistant fraction by $14.39$ percentage points and increased the model-defined cure rate by $15.33$ percentage points relative to random prescribing. Its overall frequency of the Drug 2 action was higher than under random prescribing, so the improvement was not accompanied by a reduction in Drug 2 selection. The policy shows severity-dependent prescribing behavior. It selected Drug 2 frequently for severe cases and Drug 0 frequently for mild cases, and reduced its use of the strongest drug over the course of an episode. This pattern was learned without a fixed severity-to-drug rule, although the model's severity-dependent reward and treatment-effect assumptions favored it.

Two design choices in the model push the policy toward Drug 0 for mild cases. The reward subtracts a large penalty when a mild patient receives Drug 2, and in that case the drug's kill rate is also halved. The severity-dependent pattern should therefore be read in light of these assumptions.

The simulation illustrates how a reward function can represent patient-level and ward-level objectives within a sequential decision problem. Reinforcement learning has already shown that it can learn treatment policies from clinical data, most visibly for sepsis in the intensive care unit \cite{komorowski2018clinician, nanayakkara2022sepsis, choi2024sepsis, wu2023sepsis, drudi2024icu}, and more broadly for dynamic treatment regimes and diagnostic pathways \cite{abebe2024dtr, muyama2024diagnostic, zhao2024review}. Our setting differs in what the agent must internalize. In most clinical RL the reward tracks the outcome of one patient. Here the reward also carries a population cost through a ward-level resistance term, a stylized version of the patient and ward trade-off that makes stewardship hard in practice \cite{porco2012commons, colman2019game}. Recent work has begun to move RL in this direction, by encoding clinical guidelines into the reward \cite{wang2024antibiotic, denhengst2024ventilation} and by sequencing drugs to steer resistance evolution \cite{weaver2024resistance}. Our contribution is to close the loop inside a ward whose resistance responds to the agent, with a multi-objective reward that makes the training objective explicit.

At the highest tested penalty, the policy nearly abandons the broad-spectrum drug, much like the Drug 2 restriction baseline, and its mean resistant fraction is higher, not lower, than at the main penalty setting. Stronger restriction therefore did not improve both evaluated outcomes. One possible explanation is that severe infections denied an effective drug clear slowly, so these patients stay infectious for longer, which drives up secondary transmission and environmental seeding. The resistant subpopulation carries a fitness cost and grows more slowly without the drug, but a longer infectious period gives it more time to spread \cite{andersson2010cost, czuppon2023stochastic, davies2019withinhost}. This remains an interpretation of the simulation results. Because each penalty setting was represented by a single fixed trained policy, these comparisons do not quantify variability across independent training runs.

Classical optimal control has long been used to time and size epidemic interventions \cite{srivastava2024seiqr, buelta2024control}. Those methods are powerful when the dynamics can be written in closed form, while reinforcement learning is useful when the policy has to be discovered from interaction with a complex, stochastic environment. The two are complementary, and the Pareto view we use to compare policies is a natural meeting point \cite{reymond2022pareto}. Our own earlier work supports the same theme from the side of collective behavior, where learning agents can raise group welfare in a social dilemma and can reshape interaction structure toward cooperation \cite{lee2025granular, lee2025enhancing, weng2026qlearning}. Related work from our group on coupled epidemic and behavioral dynamics examines how such systems respond to changing control strength \cite{lee2025phase}. The present study applies these ideas to a simulated prescribing setting, and it follows directly from our spatial multi-agent model of epidemic control \cite{liu2027spatial}.

The model has not been calibrated or validated against hospital data, so it is a testbed for stewardship ideas rather than a bedside controller. It is well suited to rapid screening of candidate reward structures and intervention rules, and to building intuition about where adaptive policy helps and where a simple rule is enough. Moving it toward practice would require calibration to real surveillance and prescribing data, a richer microbiological and pharmacokinetic layer, and the kind of safety-aware medication management that recent clinical RL has begun to address \cite{kim2026nephrotoxic}. Models that read resistance from hospital records \cite{hernandez2023amr} and that flag nosocomial infection early \cite{samadani2023vap} are natural inputs to such a system. Such an approach would require reliable information about patients and local resistance.

Several limitations shape how far these results should be read. First, this is a pure simulation with no real patient data, so the numbers describe the model and not any particular hospital. Second, a lower ward-averaged resistant fraction may reflect changes in patient-level resistant fractions, changes in the number of slots carrying no bacteria, or both. It does not necessarily indicate a reduction in total resistant bacterial burden. Third, the agent uses one shared policy, trained on rewards from three prescriber profiles that it cannot observe, so prescriber heterogeneity is present in the reward but not expressed in behavior. A formulation with prescriber identity in the state and decentralized actors is the natural next step, and multi-agent actor-critic methods are built for it \cite{lowe2017maddpg}. Fourth, the within-host model is deliberately simple, with three drugs and without an explicit dynamic immune-response model or comorbidities, which keeps the stewardship trade-off clean but leaves out real clinical complexity. The low frequency of Drug 1 use depends on its assigned cost and kill rates, and should not be generalized to intermediate-spectrum antibiotics in clinical practice. In addition, a patient whose load falls into the $10^3$ to $10^4$ range stops evolving unless a transmission event lifts it again, which is a simplification of the within-host dynamics. Fifth, spatial heterogeneity in resistance was small in our ward, so the spatial channel may matter more in layouts with weaker environmental mixing or more localized contact, which we leave to future work. Sixth, the reported evaluation variability does not capture variation across independently trained policies, and the training runs did not use a fixed global random seed; the saved checkpoints can still be re-evaluated with fixed evaluation seeds. The agent also observes and acts on every patient-slot each day, including uninfected ones, so it behaves more like a proactive surveillance and intervention system than a reactive prescriber that responds only to diagnosed infections, and the action-frequency summaries combine decisions for infected and uninfected slots. Finally, the policy map groups action outputs across all patient-slots into ward-resistance categories, with the high category set at a 35\% threshold. This grouping is for visualization only, and the headline outcomes in Table~\ref{tab:stewardship_outcomes} and the sensitivity analysis do not depend on it.

\section{Conclusion}
\label{sec:conclusion}

We presented a virtual hospital ward in which a reinforcement learning agent learns antibiotic prescribing from a multi-objective reward. The agent learns a severity-dependent, tiered prescribing pattern without a fixed severity-to-drug rule, and relative to random prescribing it lowers the ward-averaged resistant fraction while raising the model-defined cure rate. Among policies trained at different resistance penalties, the one at the highest penalty had a higher resistant fraction than the others. The multi-objective reward makes the training objective explicit, and the agent is not dominated by either rule-based baseline on the resistance and cure trade-off. We offer the framework as a simulation testbed for studying stewardship trade-offs. Calibration to real data and a genuinely multi-agent, heterogeneous extension are the clear next steps.

\section*{Acknowledgements}

This research was supported in part by the National Science and Technology Council of Taiwan under the grants NSTC 113-2221-E-011-150 and NSTC 114-2221-E-011-086.

\section*{Code and data availability}

The simulation code and the trained model checkpoints are available at \url{https://github.com/dukee1213/amr.git}.

\bibliographystyle{elsarticle-num}
\bibliography{AMR_references} 

\clearpage
\appendix
\section{Default model parameters}
\label{app:params}

Tables~\ref{tab:default_parameters}, \ref{tab:within_host_parameters} and~\ref{tab:reward_parameters} list the default parameters used across all experiments. These values are heuristic starting points, chosen to expose the stewardship trade-off rather than calibrated to a specific pathogen or hospital, and Appendix B probes how the main outcome responds to perturbing the transmission and mutation parameters. The severity-dependent prescribing terms described in the Methods are applied as logic in the simulator and are not scalar entries in these tables.

\begin{table}[htbp]
\centering
\small
\caption{Default ward and transmission parameters.}
\label{tab:default_parameters}
\begin{tabularx}{\linewidth}{l >{\hsize=0.6\hsize}X l >{\hsize=1.4\hsize}X}
\toprule
Symbol & Parameter & Default value & Description \\
\midrule
\multicolumn{4}{@{}l}{\textbf{Ward geometry and episode horizon}} \\
$L$ & Ward grid dimension & $50.0$ m & Length and width of the 2D spatial hospital ward \\
$N$ & Ward capacity & $200$ & Number of patients in the closed cohort \\
$T$ & Episode duration & $30$ days & Total discrete timesteps simulated per episode \\
\midrule
\multicolumn{4}{@{}l}{\textbf{Spatial transmission and shedding}} \\
$\beta_{\mathrm{env}}$ & Environmental transmission & $0.05$ & Transmission coefficient via contaminated ward surfaces \\
$\beta_{\mathrm{contact}}$ & Contact transmission & $0.15$ & Direct interpersonal transmission coefficient ($d < 2.0$ m) \\
$\omega_{\mathrm{env}}$ & Environmental shedding & $10^{-4}$ & Fraction of resistant bacterial load shed to the local grid \\
$h$ & Kernel bandwidth & $2.0$ m & Standard deviation of the spatial Gaussian shedding kernel \\
\bottomrule
\end{tabularx}
\end{table}

\begin{table}[htbp]
\centering
\small
\caption{Default within-host and pharmacodynamic parameters.}
\label{tab:within_host_parameters}
\begin{tabularx}{\linewidth}{l >{\hsize=0.6\hsize}X l >{\hsize=1.4\hsize}X}
\toprule
Symbol & Parameter & Default value & Description \\
\midrule
\multicolumn{4}{@{}l}{\textbf{Within-host microbial dynamics}} \\
$K$ & Carrying capacity & $10^9$ CFU & Maximum within-host bacterial population per patient \\
$r_S$ & Susceptible growth rate & $0.6$ day$^{-1}$ & Intrinsic logistic growth rate of the susceptible strain \\
$r_R$ & Resistant growth rate & $0.55$ day$^{-1}$ & Intrinsic growth rate of the resistant strain (fitness cost) \\
$\mu_S, \mu_R$ & Natural clearance rate & $0.1$ day$^{-1}$ & Baseline host clearance rate without antibiotics \\
$m$ & Mutation rate & $10^{-6}$ day$^{-1}$ & Transition rate from susceptible to resistant strain \\
\midrule
\multicolumn{4}{@{}l}{\textbf{Antibiotic action space and pharmacodynamics}} \\
$\phi_0, \phi_1, \phi_2$ & Baseline kill rates & $2.2, 2.4, 2.5$ day$^{-1}$ & Kill rates against the susceptible strain for Drugs 0, 1, and 2 \\
$\alpha_0, \alpha_1, \alpha_2$ & Residual efficacy factors & $0.15, 0.08, 0.20$ & Kill-rate multiplier against the resistant strain \\
$c_0, c_1, c_2$ & Direct antibiotic costs & $0.1, 1.0, 3.0$ & Base economic cost incurred per drug administration \\
\bottomrule
\end{tabularx}
\end{table}

\begin{table}[htbp]
\centering
\small
\caption{Multi-objective reward weights and scaling. Each weight vector is $(\omega_\alpha, \omega_\beta, \omega_\gamma, \omega_\delta)$ for treatment efficacy, resistance prevention, cost, and adverse outcomes.}
\label{tab:reward_parameters}
\begin{tabularx}{\linewidth}{l >{\hsize=0.7\hsize}X l >{\hsize=1.3\hsize}X}
\toprule
Symbol & Parameter & Default value & Description \\
\midrule
$\boldsymbol{\omega}_{\mathrm{guide}}$ & Guideline reward weights & $(1.9, 1.8, 0.3, 0.1)$ & Applied to 50\% of prescribers \\
$\boldsymbol{\omega}_{\mathrm{cons}}$ & Conservative reward weights & $(1.8, 2.0, 0.3, 0.1)$ & Applied to 30\% of prescribers \\
$\boldsymbol{\omega}_{\mathrm{agg}}$ & Aggressive reward weights & $(2.0, 0.2, 0.2, 0.2)$ & Applied to 20\% of prescribers \\
$\omega$ & Ward resistance weight & $20.0$ & Scaling weight for the quadratic ward-level resistance penalty \\
$\eta$ & Resistance penalty scalar & $0.4$ & Effective global multiplier ($8.0/20.0$) on the resistance penalty \\
\bottomrule
\end{tabularx}
\end{table}

\section{Sensitivity analysis}

Supplementary Table~\ref{tab:sensitivity} reports a sensitivity analysis over 10 evaluation episodes per condition. The outcomes stay broadly consistent with the main results in Table~\ref{tab:stewardship_outcomes}. Raising either transmission parameter increased the terminal resistant fraction, with $\beta_{env}$ ($+20\%$) giving the largest single change. Lowering the parameters reduced it, and the contact channel showed a slightly larger change in that direction. Within the tested $\pm20\%$ perturbations, the terminal resistant fraction changed less under mutation-rate perturbations than under transmission-rate perturbations. With 10 evaluation episodes per condition, estimates of both the mean and the standard deviation are imprecise.

\begin{table}[h]
\caption{Supplementary sensitivity analysis of key epidemiological parameters ($N=10$ evaluation episodes per condition).}
\label{tab:sensitivity}
\begin{center}
\begin{tabular}{|l|c|c|}
\hline
\rule[-1ex]{0pt}{3.5ex} \textbf{Perturbation condition} & \textbf{Terminal resistant fraction (\%)} & \textbf{Model-defined cure rate (\%)} \\
\hline
\rule[-1ex]{0pt}{3.5ex} Baseline Control & $46.88 \pm 5.40$ & $47.60 \pm 7.16$ \\
\hline
\rule[-1ex]{0pt}{3.5ex} $\beta_{env}$ ($+20\%$) & $49.70 \pm 4.88$ & $47.39 \pm 7.34$ \\
\hline
\rule[-1ex]{0pt}{3.5ex} $\beta_{env}$ ($-20\%$) & $46.47 \pm 5.42$ & $48.12 \pm 7.13$ \\
\hline
\rule[-1ex]{0pt}{3.5ex} $\beta_{contact}$ ($+20\%$) & $49.10 \pm 5.16$ & $47.19 \pm 8.30$ \\
\hline
\rule[-1ex]{0pt}{3.5ex} $\beta_{contact}$ ($-20\%$) & $45.95 \pm 5.31$ & $51.09 \pm 8.98$ \\
\hline
\rule[-1ex]{0pt}{3.5ex} $m$ ($+20\%$) & $47.10 \pm 6.12$ & $45.08 \pm 5.40$ \\
\hline
\rule[-1ex]{0pt}{3.5ex} $m$ ($-20\%$) & $46.88 \pm 5.40$ & $47.43 \pm 6.50$ \\
\hline
\end{tabular}
\end{center}
\end{table}

\end{document}